\documentclass{aa}

\usepackage{graphicx}

\begin{document}

\title{The possible orbital period of the nova V1493 Aquilae}

\author{A.~Dobrotka \inst {1}, M.~Friedjung \inst {2}, A.~Retter \inst {3,4}, L.~Hric \inst {5}, R.~Novak \inst {6}}

\offprints{A.~Dobrotka}

\institute{Departement of Physics,
           Faculty of Materials Science and Technology,
           Slovak University of Technology in Bratislava,
           Paul\'{\i}nska 16,
           91724 Trnava,
           The Slovak Republic,
           (e-mail:~andrej.dobrotka@stuba.sk)
\and
           Institut d'Astrophysique de Paris,
	   98 bis, boulevard Arago,
	   75014 Paris,
	   France
	   (e-mail:~fried@iap.fr)
\and       
           Departement of Astronomy and Astrophysics,
           Penn State University,
           525 Davey Lab, University Park,
           PA, 16802-6305,
           USA
           (e-mail:~retter@astro.psu.edu)
\and
           School of Physics,
           University of Sydney, 2006,
           Australia
\and
           Astronomical Institute of the Slovak Academy of Sciences,
           05960 Tatransk\'{a} Lomnica,
           The Slovak Republic,
           (e-mail:~hric@ta3.sk)
\and
           Nicholas Copernicus Observatory and Planetarium in Brno,
           Krav\'i Hora 2,
           61600 Brno,
           Czech Republic,
           (e-mail:~novak@hvezdarna.cz)
           }

\date{Received / Accepted}

\abstract
   {}
   {Period analysis of CCD photometry of V1493~Aql (Nova Aql 1999 no. 1) performed during 12 nights through $I$ and $R$ filters a few weeks after maximum is presented.}
   {The PDM method for period analysis (Stellingwerf 1978) is used.}
   {The photometric data is modulated with a period of 0.156 $\pm$ 0.001 d. Following the sinusoidal shape of the phased light curve, we interpret this periodicity as possibly orbital in nature and that is consistent with a cataclysmic variable above the period gap.}
   {}

\maketitle \markboth{A.~Dobrotka et al.: The possible orbital period of the nova V1493 Aquilae}{}

\section{Introduction}

Nova Aql~1999 no. 1 (V1493~Aql) was discovered on July $13^{th}$ 1999 (Nakano and Tago 1999). The nova reached a maximum of V $\sim$ 8.8 and faded to V $\sim$ 15.5 four months later. The nova underwent a secondary brightening with a maximum of V $\sim$ 11.4 approximately 45 days after maximum. Moro et al. (1999) could not find any indication of dwarf nova outbursts on 40 Asiago plates spanning 1962-1982 to a limiting magnitude of 17.5. Tomov et al. (1999) studied spectral observations made a few days after discovery and found an expansion velocity of $\sim 1700$ km~s$^{-1}$, the same value as derived by Ayani et al. (1999). They announced strong emission of Fe~II multiplets and the absence of strong forbidden lines. No line showed a convincing P~Cyg profile. Arkhipova et al. (2002) carried out spectroscopic observations of the nova from July $17^{th}$ 1999 until November $5^{th}$ 1999. They derived the distance to be 4.2 kpc. In the spectrum all lines were broad (FWHM $\sim$ 3400 km~s$^{-1}$). Their July spectra exhibited Balmer hydrogen lines, lines of ionized iron, N II 5680 and possibly He I 5876. On August $4^{th}$ the iron lines disappeared and the nebular lines began to grow ([O II] 7320+7330, [N II] 5755). On September $16^{th}$ the nova was at the nebular phase. The He I 6678, 7065, [O I] 6300, 6364 lines appeared. A 0.8 -- 2.5 $\mu$m spectroscopic study was presented by Lynch et al. (2000), who found very broad lines (FWHM $\sim 4000$ km~s$^{-1}$). They did not detect the presence of a dust thermal emission. Venturini et al. (2004) analysed the same spectroscopy and announced that the spectrum was populated by blended low-excitation lines (H I Brackett and Paschen and O I). The He II line started to emerge. They concluded that the nova experienced a second period of mass loss. They calculated the distance to be 25.82 $\pm$ 1.81 kpc.

Previous studies of the photometry and the periodic modulation ($\sim 0.156 \pm 0.001$ days) were presented by Novak, Retter \& Lipkin  (1999) and Dobrotka et al. (2005). In this work we give a detailed analysis and discussion of these photometric data. After the presentation of the observations in Section II we describe the period analysis of our data (Section III). In the final section (Section IV) we discuss our results.

\section{Observations}

Most of our photometry was performed with an ST-7 CCD camera using an $R$ filter at the Nicholas Copernicus Observatory and Planetarium in Brno, Czech Republic. We observed the nova during 11 nights from July $20^{th}$ 1999 to August $22^{th}$ 1999 with the 0.4-m Newtonian telescope. For the reduction of the data we used the aperture photometry package Muniphot based on Stetson's DaoPhot (from MIDAS).

A single run was conducted with the 1.0-m telescope at the Wise~Observatory (WO) Israel, on Aug 1$^{st}$ 1999, using the Tektronix 1K back-illuminated CCD (see Kaspi et al. 1995 for details of the telescope and instrumentation). Observations were conducted through an $I$ filter, with typical exposure times of 30~sec. Bias and flat-field correction of the images were performed in the standard fashion. Photometric measurements of the corrected images were carried out using the NOAO IRAF\footnote{IRAF (Image Reduction and Analysis Facility) is distributed by the National Optical Astronomy Observatories, which are operated by AURA, Inc., under cooperative agreement with the National Science Foundation.} {\sc{daophot}} package (Stetson 1987). Instrumental magnitudes of V1493 Aql as well as of 44 reference stars were measured in each frame. A set of internally consistent magnitudes of the object was obtained using the WO reduction program {\sc{daostat}} (Netzer et al. 1996).

\begin{figure}
  \vspace{0mm}
  \hspace{0mm}
  \resizebox{8.7cm}{!}{\includegraphics{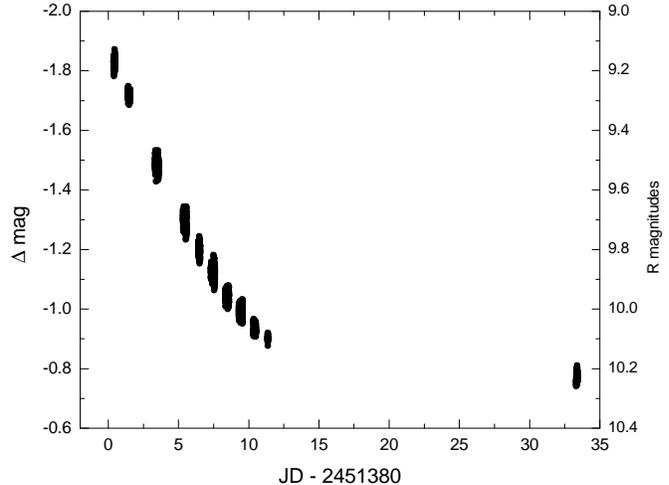}}
  \caption[ ]{The $R$ band light curve of our data. The calibration of the right hand axis was made with VSNET $R$ observations.}
  \label{Rcourve}
\end{figure}

Table 1 lists the photometric observations. The light curve is presented in Fig. \ref{Rcourve}. The data are available online from the electronic website of the journal \footnote{The data are only available in electronic form at the CDS via anonymous ftp to cdsarc.u-strasbg.fr (130.79.128.5) or via http://cdsweb.u-strasbg.fr/cgi-bin/qcat?J/A+A/}.

\begin{table}
\caption{The observational log. There are runs from two
observatories: Brno (B), Wise Observatory (WO). HJD is -2451000 days, t is the duration of the observation in hours and N is number of frames.}
\begin{center}
\begin{tabular}{rcccrcccc}
\hline
\hline
 Obs. & Date & HJD (start) & t [h] & N \\
\hline
 B & 20.7.1999 & 380.3999 & 1.96 & 191 \\
 B & 21.7.1999 & 381.3984 & 3.96 & 368 \\
 B & 23.7.1999 & 383.3478 & 5.79 & 552 \\
 B & 25.7.1999 & 385.3422 & 5.68 & 586 \\
 B & 26.7.1999 & 386.4170 & 3.17 & 327 \\
 B & 27.7.1999 & 387.3247 & 6.56 & 613 \\
 B & 28.7.1999 & 388.3362 & 6.32 & 579 \\
 B & 29.7.1999 & 389.3144 & 6.49 & 826 \\
 B & 30.7.1999 & 390.3406 & 4.33 & 306 \\
 B & 31.7.1999 & 391.3613 & 0.41 & 39 \\
 WO & 01.8.1999 & 392.4814 & 2.18 & 171 \\
 B & 22.8.1999 & 413.3265 & 2.58 & 192 \\
\hline
\end{tabular}
\end{center}
\end{table}
\begin{figure}
  \vspace{0mm}
  \hspace{0mm}
  \resizebox{7cm}{!}{\includegraphics{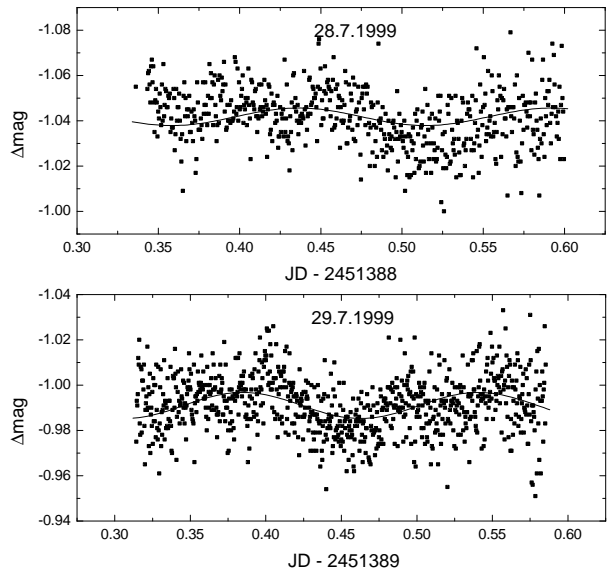}}
  \caption[ ]{Light curves of two of our best nights in 1999 obtained using an $R$ filter. The upper panel shows the points of July 28$^{th}$ 1999 and the lower panel -- July 29$^{th}$ 1999. The line is the sinusoidal fit to the data using the parameters derived in this paper.}
  \label{2nights}
\end{figure}

\section{Period analysis}

\begin{figure}
  \vspace{0mm}
  \hspace{0mm}
  \resizebox{8cm}{!}{\includegraphics{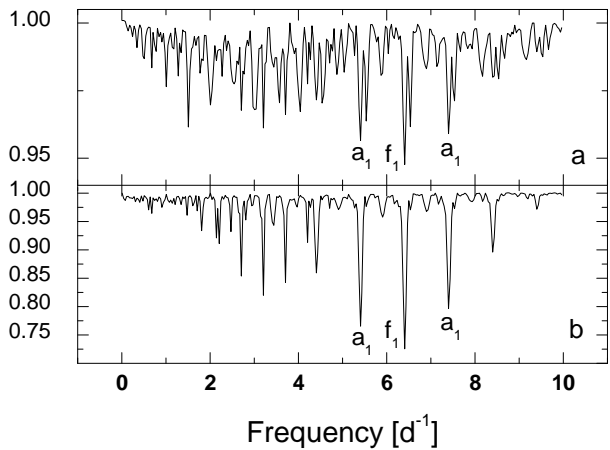}}
  \caption[ ]{Power spectra of all data. The orbital frequency (f$_{\rm 1}$) with aliases (a$_{\rm 1}$) is shown, a -- raw data, b -- power spectrum of the synthetic curve.}
  \label{poweranalys}
\end{figure}
\begin{figure}
  \vspace{0mm}
  \hspace{0mm}
  \resizebox{8cm}{!}{\includegraphics{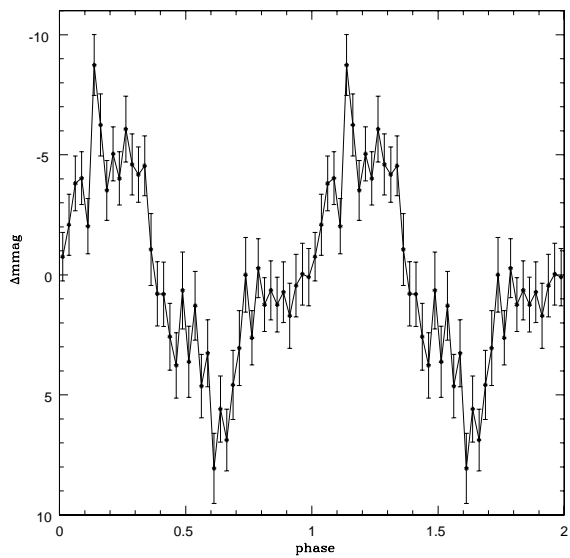}}
  \caption[ ]{The $R$ band light curve folded on the 3.7 h period and binned into 40 equal bins. The error of the mean is shown.}
  \label{foldedcourve}
\end{figure}
Our 12 runs taken in $I$ and $R$ filters were used for period analysis. Fig. \ref{2nights} presents two examples of nightly runs -- July $28^{th}$ and July $29^{th}$. The total observing time was $\sim$50 hours and the mean nightly run was $\sim$4.1 hours. The total number of frames was 4750 and approximately 396 frames per a single run. The mean error of the observational data was $\sigma = 0.01$ mag. The data were detrended by subtracting the mean values from each night. The best-fitted ephemeris for the periodicity is:
\vskip5mm
$T_{min} = HJD~2451380.415~(\pm 0.001) +  0.156 E~(\pm 0.001)$
\vskip5mm
For the period analysis we used the PDM method (Stellingwerf 1978). The power spectra of all data are shown in Fig. \ref{poweranalys}a. The highest peak (f$_{\rm 1}$) is at 6.41 d$^{-1}$ (P = $0.156 \pm 0.001$ day). The peaks around it are identified as aliases (a$_{\rm 1}$). To check whether our identification of period and aliases is real, we carried out the following tests.

First, we made synthetic power spectrum (Fig. \ref{poweranalys}b) of a sinusoid with the frequency 6.41 d$^{-1}$ sampled and noised as the data. This test showed that all typical features in Fig. \ref{poweranalys}a are simply explained by the periodicity found in the data. We then performed period analysis of the first 6 nights and the last 6 nights separately. The period appeared in both power spectra and this supports the presence of the periodicity.

The $R$ band light curve folded on the 3.7 h period is depicted in Fig. \ref{foldedcourve}. The shape of the curve is very shallow with a peak-to-peak amplitude $\sim 15$ mmag.

\section{Discussion}

We found a period modulation of 3.7 hours in all data. The phased light curve shows typical behaviour of an eclipse. This could be a strong argument that the detected modulation is the orbital period of the binary system. However, as we shall see later, an eclipse interpretation leads to difficulties. Hence Nova V1493~Aql with its newly discovered orbital period may belong to the markedly numerous systems with orbital periods above the period gap and is then a member of the largest groups in the interval between 3 and 4 hours.

The folded light curve suggests that the periodic variations were already observed only 5 days after the visual maximum at JD 2451375 estimated by Venturini et al. (2004). The fairly constant period could be orbital; to see orbital variations so early is suprising, and might suggest that an elongated object, associated with the central binary (possibly a common envelope binary), was weakly visible below a possibly non-spherically symmertric almost optically thick wind. If we had seen a non-expanding atmosphere of an object with a radius near 10$^{12}$~cm, rotating with the observed period, most of the outer layers would have been violently rotationally unstable. We can justify such a radius, as it should be of the same order as that from the fits of optically thick wind calculations by Short et al. (2001), giving a photospheric radius at 5000 $\AA$ of nova V1974 Cyg of not less than 10$^{12}$~cm up to 2-3 weeks after the maximum radius. That nova faded 3 magnitudes in V from maximum in 42 days  (Chochol et al. 1993), while the faster V1493 Aql faded the same amount in 23 days (Venturini et al. 2004). Eclipses, if real, would be even more surprising. The eclipse-like feature lasted about 0.2 of the period with an amplitude of about 0.5\%.
A calibrated spectrum in our possession suggests that rather less than half the luminosity in $R$ on 1999 July 24$^{th}$ came from the continuum. The region of origin of most of the line emission, including the very strong H$\alpha$ emission, is model dependent and much might come from relatively dense outer optically thin parts of the envelope rather than from near the photosphere (for instance see model of Friedjung 1987), making real eclipses larger in the continuum than in integrated light in $R$. We can use the expressions of Warner (1995) and Drilling and Landolt (2000) to find the characteristics of the secondary and see if eclipses were possible. We find that it is difficult to produce eclipses, even with very favourable assumptions, unless V1493 Aql was not a classical nova and had a luminosity well below the Eddington limit near maximum. In any case more observations are needed of this now faint, probably nearly quiescent system, to elucidate its nature.

\section{Summary}

We found a periodic signal of 3.7 hours in 50 hours of $R$ and $I$ photometry obtained in the first month of outburst of V1493 Aql. The shape of the light curve and the duration of the observed period is indicitive of an orbital period above the period gap. However, more observations are required to confirm if this early periodicity is in fact orbital in nature.

\acknowledgements

This work was supported by the Slovak Academy of Sciences Grant No. 4015/4 and partly by the Science and Technology Assistance Agency under the contract No. 51-000802. We thank Ohad Shemmer and Yiftah Lipkin for valuable comments and for obtaining part of the photometric and spectroscopic observations. AR was partially supported by a research associate fellowship from Penn State University. We are grateful to the VSNET observers. We acknowledge the anonymous referee for valuable comments.

\end{document}